\documentclass[aps,prl,twocolumn,groupedaddress,showpacs]{revtex4}
\def\beq{\begin{equation}}
\def\eeq{\end{equation}}
\def\beqa{\begin{eqnarray}}
\def\eeqa{\end{eqnarray}}

\def\bdi{\begin{displaymath}}
\def\edi{\end{displaymath}} 

\usepackage{graphicx,amssymb,amsmath,float}

\begin{document}

\title{Universality classes for depinning transitions of two heteropolymers}
\author{D.~Suppa}
\affiliation{Theoretical Physics, University of Oxford, 1 Keble Road,
Oxford, OX1 3NP, UK}
\affiliation{Dipartimento di Fisica Universit\`a  di Roma ``La Sapienza'', Piazzale Aldo Moro 5,
I-00185 Roma, It}
\date{\today}

\begin{abstract}
We study a solid-on-solid model for depinning transitions of two directed heteropolymers
with an effective long range interaction decaying as the inverse square of their distance. 
Exact calculations of the localization length and specific heat non-universal critical exponents 
indicate that new universality classes govern the depinning transitions, different from those found 
for interacting homopolymers. Disorder driven change in the order of the phase transition and in the sign 
of the specific heat exponent are predicted.
   
\end{abstract}



\maketitle
\begin{figure}
\includegraphics[width=7cm]{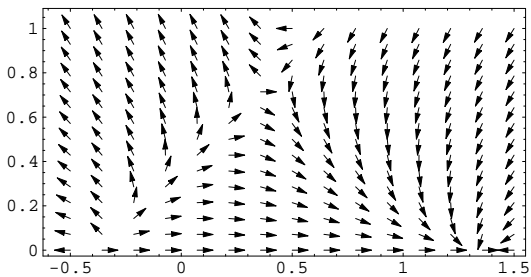}
\caption{A cartoon of the RG flow in $d=1$ and $g=0.5$. The disorder variance is along the
$y$-axis and $u$ is along the $x$-axis.} 
\label{RGcartoon1}
\end{figure}
\begin{figure}
\includegraphics[width=7cm]{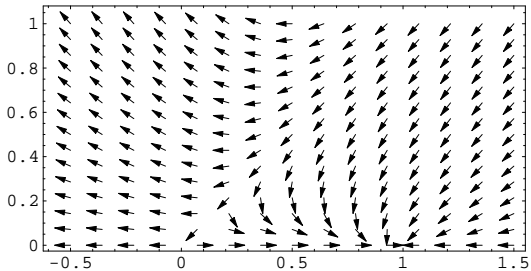}
\caption{A cartoon of the RG flow in $d=1$ and $g=0$. The disorder variance is along the
$y$-axis and $u$ is along the $x$-axis.} 
\label{RGcartoon2}
\end{figure}

The unbinding of two directed polymers \cite{somen,straley,lassig,kolomeisky,mukamel} is an important 
example of a class of surface depinning phenomena, namely 
surface growth \cite{parisi}, denaturation of double-stranded DNA (dsDNA) molecules 
\cite{poland,wartell}, wetting on a substrate \cite{douglas,forgacs,derrida}, depinning of flux lines from a columnar 
defect in a type II superconductor \cite{superconductor}, etc.
Our understanding of these surface phenomena is mainly based on the analysis of a class of models using the
solid-on-solid (SOS) approximation which accounts for the crucial surface fluctuations and ignores irrelevant
bulk degrees of freedom.
A challenging topic in these systems is to understand whether the presence of randomness is relevant at the transition
and thus modifies the critical behavior \cite{fisher}. The purpose of this paper is to discuss this question
in a simplified model, where the original problem of two interacting heteropolymers is formulated as that
of an interface near a one-dimensional random surface.
In the absence of long range interactions, a similar problem has been addressed in the past in the context of 
wetting \cite{douglas,forgacs,derrida} and, in two dimensions, it is known that the transition still exists and the 
disorder is marginal in a renormalization group (RG) sense, thus requiring a perturbative analysis to establish 
whether it will move the system away from the pure case fixed point or not \cite{forgacs,derrida,kallabis,hwa}.
A perturbative analysis to all orders in the strength of the disorder was carried out in \cite{forgacs} where
only a subleading order correction to the singular part of the free energy at the transition of the pure case
was found. However, a perturbative expansion in an RG form \cite{derrida} showed that disorder 
shifts the transition temperature and also modifies the critical behavior, though it was not possible to predict 
the new critical exponents because the RG flux flows toward strong-disorder regions. 
Such results were confirmed by a numerical analysis in \cite{hwa}, while the issue of studying the strongly 
disordered regime was addressed in \cite{tang,mukamel}.   
However, the question whether the introduction of disorder gives rise to an RG flow toward strong-coupling regime 
or a new fixed point exists for the random system in the perturbative regime is still debated, as well as the 
problem of determining the order of the phase transition \cite{mukamel,tang,azbel,causo,kafri,stella}, both in the
presence and in the absence of long range interactions.\\
We consider here an SOS model of two directed heteropolymers with a long range tail interaction 
decaying as the inverse square of their distance \cite{somen,straley}. 
This is required to model, e.g., the effective interaction of steps on a vicinal surface
\cite{lassig,Song}, and similar long range interactions occur in the problem of dsDNA denaturation, where
they are  meant to mimic a space-dependent stiffness, due to the fact that the double-stranded conformation is 
significantly more rigid than the single one \cite{dauxois,hwa,gotoh,stella}. 
Modelling the original system as an interface near a wall and in the absence of disorder,
an exact solution of the critical exponents \cite{somen,straley} allows one to predict a nontrivial 
first order phase transition
above an upper critical dimension, second order phase transitions with non-universal exponents, or 
Kosterlitz-Thouless-like behavior upon changing the amplitude of the repulsive tail.
When disorder is added along the wall, we  find new universality classes for the depinning transitions,
compute exactly the localization length and specific heat non-universal critical exponents and show that disorder 
can induce a change in the order of the phase transition as well as
in the behavior of the specific heat at criticality. We emphasize the paradigmatic nature of the
present model for surface depinning transitions in systems where the disorder is localized on the wall, 
in view of its exact solvability.\\
We begin by considering a linear object directed along, say, the z axis but fluctuating in the transverse
$d$-dimensional $\vec x$ space where it is subject to an isotropic potential ${\cal V}(|\vec x|,z)$. 
The corresponding continuum Hamiltonian is $H=\int dz\left\{\frac{m}{2}\left(\frac{d \vec x}{dz}
\right)^2 +{\cal V}(|\vec x|,z)\right\}$, where $m$ is the line stiffness and $\vec x (z)$ is the relative
coordinate of the two original directed heteropolymers. 
The isotropic potential ${\cal V}(|\vec x|,z)$ consists of a short range pinning term and of a long range 
power law repulsive tail
${\cal V}(|\vec x|,z)=v^0(z)\; \delta^{d}_{\Lambda}(\vec x)+\frac{V^0_S}{|\vec x|^s} 
\;(2 \pi)^d \theta(|\vec{x}|-\Lambda^{-1})$,
where $\theta(x)$ is the Heaviside step function, $\delta^{d}_{\Lambda}({\vec x})=\Lambda^{d}
\frac{d}{K_d}\;\theta(\Lambda^{-1} -|\vec x |)$, and $K_d=S_d/(2 \pi)^d$ with $S_d$ the 
surface area of a d-dimensional unit sphere. $\Lambda$ is a momentum cutoff set by the 
short-range part of the potential. Although in this paper we will restrict our analysis  
to the case $d=1$ and $s=2$, we keep $d$ and $s$ general in the derivation of our RG equations. 
The partition function can be evaluated by mapping the problem into a quantum-mechanics
problem in d dimension through the transfer operator technique \cite{douglas,Chalker,Chui,Lipowsky}. 
In this way the most singular part $F$ of the line free-energy is given, in the thermodynamic limit, 
by the lowest energy eigenvalue $E_0$ of the Schr$\ddot{o}$dinger equation for the wavefunction of the
corresponding quantum-mechanics problem.
Two length scales (characterizing the correlations along the polymer and in the transverse plane, 
respectively) are involved in the depinning transition: $\xi_{\parallel}$ and $\xi_{\perp}$. 
They are given in terms of the temperature $T$ and energy $E_0$ by the scaling relations 
$\xi_{\parallel}=T/|E_0|$ and $E_0=-DT^2/m \xi_{\perp}^2$ (where D is a constant), and are thus related:
$\xi_{\perp}^2=(DT/m)\xi_{\parallel}$ \cite{fisher}. The phase transition can thus be studied in terms
of the length scale $\xi_{\perp}$ alone.\\
We define $v^0 (z)$ as the sum of a constant term plus a quenched degree of freedom along the coordinate z:
$v^0 (z)= \tilde{v}^0+\delta \tau(z)$.
To the quenched disorder we associate a Gaussian probability distribution
$P([\delta \tau])={\cal D}[\delta \tau] \exp{\left( - \frac{1}{2 \Delta^0} 
\int_{-\infty}^{+\infty}\delta \tau^2(z)dz
\right)}$with variance $\Delta^0 <<1$.
We treat the general problem of studying the whole set of 
RG equations in the presence of quenched disorder by using the replica approach. 
After Gaussian integration over $\delta \tau$, the replicated partition function is
$Z_{n}=\int {\cal D}[\vec{x}^{a}] e^{-\int dz \; {\cal H}_n}$, with
\begin{eqnarray}
\label{repl_ham}
&{\cal H}_n& = \sum_{a=1}^{n} \Big[ 
\frac{m}{2}\left(\frac{d \vec x ^a}{dz}
\right)^2 + \frac{V^0_S}{|\vec x ^a|^s}(2 \pi)^d \theta(|\vec{x}^a|-\Lambda^{-1})\\ \nonumber
&+&\left(\tilde{v}^0 -\frac{\Delta^0 \Lambda^d d}{2 K_d} \right) \delta^{d}_{\Lambda}(\vec x ^a) 
-\frac{\Delta^0 \Lambda^d d}{2 K_d}\sum_{b\neq a}\delta^{d}_{\Lambda}(\vec x ^a)
{\cal N}_{a,b}\Big]
\end{eqnarray}
where the interactions between replicas give rise to a renormalization
of the contact energy plus an inter replicas term and we have defined
${\cal N}_{a,b}=\int_{0}^{\Lambda^{-1}} d^{d} \vec{x}^{b} 
\delta^{d}(\vec{x}^{a} - \vec{x}^{b})
=0\;{\mbox{or}}\;1\;{\mbox{for}}\;a\neq b $, depending on whether the 
positions of the replicas $a,b$ coincide inside the stripe
$[0,\Lambda^{-1}]$ or not.
We thus implement a momentum-shell RG approach.
In the momentum space the interaction potential is ${\cal V}_n[\{\vec{k}^{a}\}]=
\sum_{a=1}^{n}{\cal V}_{a}(|\vec{k}^{a}|)$, where
\beq\label{V_ab}
{\cal V}_{a}(|\vec{k}^{a}|)=
\Big[ \left(w+ \frac{V^0_S A(d,s) |\vec{k}^a|^{s-d}}{d-s} \right) 
-\sum_{b \neq a} \frac{\Delta^0 \Lambda^d d}{2 K_d} {\cal N}_{a,b} \Big]
\eeq 
and $w=V_0 -\frac{\Delta^0 \Lambda^d d}{2 K_d}$, $V_0= \tilde{v}^0-\frac{V^0_S A(d,s)
\Lambda^{s-d}}{d-s}$.
The integration over fast degrees of freedom is computed perturbatively up to the second order in 
${\cal V}_{a}(|\vec{k}^{a}|)$, using the crucial fact that disorder interactions are localized on the wall. 
It proves useful to introduce new dimensionless variables $u(l)$, $g(l)$ and 
$\hat{\Delta}(l)$ 
\begin{eqnarray}
\label{dimensionless_var1}
g(l)&=& B(\Lambda,d) A(d,s) \Lambda^{s-d} V_s(l) \\
\label{dimensionless_var2}
u(l)&=&  B(\Lambda,d)V_0(l)+ \frac{g(l)}{d-s}\\
\label{dimensionless_var3}
\hat{\Delta}(l)&=& B(\Lambda,d) \frac{\Lambda^{d} d}{2 K_d} \Delta(l)
\end{eqnarray}
where $B(\Lambda,d)=\frac{2mK_d \Lambda^{d-2}}{T^2}$.
In the limit $n \rightarrow 0$ and for $s-d<2$, as we will assume is the case in the following, 
the small-$\vec k$ behavior is governed by the following RG equations 
\begin{eqnarray}
\label{RG-1}
\frac{du(l)}{dl} & = & (2-d)u(l)-u^2(l)+g(l) -d \hat{\Delta}(l)\\
\label{RG-2}
\frac{dg(l)}{dl} & = & (2-s)g(l) \\
\label{RG-3}
\frac{d\hat{\Delta}(l)}{dl} & = &2(1-d)\hat{\Delta}(l) -2\hat{\Delta}(l)u(l)
+ \hat{\Delta}^2(l)
\end{eqnarray}  
Eq.s (\ref{RG-1})-(\ref{RG-3}) are to be solved subject to the initial conditions
obtained by evaluating eqs.(\ref{dimensionless_var1})-(\ref{dimensionless_var3})
at $l=0$, i.e. with all variables set equals to their bare values.
We stress that in the case $s=2$ and $d=1$, the initial conditions for 
$g$ and $\hat{\Delta}$ are cutoff independent.
It is easy to check that in the absence of disorder we find the same equations as in
\cite{straley}. Moreover, in the absence of the tail and in d=1, after taking 
the continuous limit $\Lambda \rightarrow +\infty$ we obtain the result that the 
disorder is marginally relevant as in \cite{derrida}.   
In the remaining of this paper we shall consider the marginal case $s=2$ only.
In this case, it follows from eq.(\ref{RG-2}) that the amplitude of the long-range potential 
is scale invariant and plays the role of a parameter in eq.(\ref{RG-1}).
Eq.s (\ref{RG-1})-(\ref{RG-3}) thus have two sets of real solutions:
the first one describes the pure system without disorder and $g \geq 0$ \cite{straley}: 
\beq\label{1-RG-sols}
\hat{\Delta}=0 \;\; ; \;\;
u_{s,u}=\frac{(2-d)\pm \sqrt{(d-2)^2+4g}}{2} \;\;
\eeq 
while the second one refers to the disordered system and is the following:
\beq\label{2-RG-sols}
\hat{\Delta}^{\star}=-d+\sqrt{(d-2)^2+4g}\; ; \;
u^{\star}=(1-d)+\frac{\hat{\Delta}^{\star}}{2} \;\;
\eeq
with $g$ satisfying the inequality $g\geq \frac{d^2-(d-2)^2}{4}=d-1$. 
Before discussing the physics behind eqs.(\ref{RG-1})-(\ref{RG-3}), let us 
discuss why they can be used for an exact determination of critical singularities
although they arise from perturbative RG. The exactness of eq.(\ref{RG-2})
is due to the fact that the thin-shell integration is analytic and therefore does not 
renormalize a singular $V_S$-type interaction, whose RG equation thus follows from dimensional 
analysis. Regarding eqs.(\ref{RG-1}) and (\ref{RG-3}), we will now 
take advantage from the equivalence between the present system and the one-particle
quantum-mechanics problem in d dimensions, and prove
that those equations correspond to the rigorous solution of the Schr$\ddot{o}$dinger equation
at large distances for the wave function of the state with zero energy. In fact,
considering a radially symmetric wave function and defined $R(r)$ as its radial part, 
the state with zero energy obeys the following equation
\begin{equation}
\label{radial-eq}
R'' + \frac{d-1}{r}R' - \frac{2mV_s}{T^2 r^s}R=0
\end{equation}
We then make the following ansatz for the function $R(r)$
\begin{equation}
\label{ansatz-eq}
R=\mbox{const}\; exp \int \frac{u(r)-\hat{\Delta}(r)}{r} dr
\end{equation}
Then, introducing (\ref{ansatz-eq}) into eq.(\ref{radial-eq}) and using (\ref{RG-3}) we
obtain for $u(r)$ the same equation as in (\ref{RG-1}), provided $l=ln(r/\alpha)$ where
$\alpha$ is a cut-off defined by the equivalence 
$2mV_s/T^2 \alpha^{s-2}=(2mK_d A/T^2 \Lambda^{2-s})V_s$. All the informations regarding
the issue of matching the long-range solution of the wave function with its short-range
part are encoded into the initial conditions.
This proves that eqs.(\ref{RG-1}),(\ref{RG-3}) give us the spatial behavior of the
wave function with $E=0$. The number of zeros of $R(r)$ coincides with the number of
lower ($E<0$) discrete levels ({\it``oscillation theorem''}). In the proximity of the phase 
transition the discrete spectrum disappears and the position of the zero of the $E=0$ wave 
function occurs at larger and larger distances from the origin. The scaling argument proposed in
\cite{straley} allows one to identify the position of this zero with the scale $\xi_{\perp}$. 
Hence we need to find a finite scale $l^\star$ at which the $R(r)$ vanishes and then study
how the corresponding localization length diverges on approaching the phase transition.\\
In eq.(\ref{1-RG-sols}), the symbols $s$ and $u$ are used for stable and unstable 
fixed points, respectively. 
$u_s$ describes the properties of the unbound phase, while $u_u$ is
the unstable fixed point where a binding-unbinding transition occurs. For 
$\sqrt{(d-2)^2+4g}<d$ only those solutions exist while the one
in (\ref{2-RG-sols}) becomes unphysical, because the parameter $\hat \Delta$ can be only positive.
In this case the depinning transition occurs at $u_u$ and there is a scale $l^{+}$ for 
which $u(l \rightarrow l^{+})\rightarrow - \infty$, indicating the formation of a bound state. 
To the corresponding diverging spatial scale $\xi_{\perp}=\Lambda^{-1}e^{l^{+}}$ it can be 
associated the exponent $\nu_{\perp}=(u_s -u_u)^{-1}$. Then, using the scaling relation 
$\alpha=2-2d\nu_{\perp}$, the specific heat exponent is also obtained.
One finds that for $u_s -u_u>2$ there is a region
of the phase space where the transition is first order with a jump of the first 
derivative of the free energy, while for $0<u_s-u_u<2$ the transition is second order 
and the critical exponents depend upon $g$ and are thus non-universal 
\cite{straley},\cite{somen}.\\
A stability analysis of the RG equation (\ref{RG-1})-(\ref{RG-3}) shows however that 
the {\it ``pure''} fixed point ($\hat \Delta=0 \; , u=u_u$) is unstable after introducing
disorder. In the presence of quenched disorder, depending on the initial conditions, 
RG flux flows either toward strong coupling regime or toward the new {\it ``random''} fixed point 
given by eq.(\ref{2-RG-sols}) which is the new locus of the depinning transition, provided
$(d-1)\leq g \leq (d+2)(3d-2)/4$. A cartoon of the RG flow is shown in Fig.\ref{RGcartoon1} 
for $d=1$ and $g=0.5$: the presence of a new unstable fixed point for the disordered system 
together with the two critical points of the pure system is apparent.
In terms of what happens into the argument of the function $R(r)$ defined in 
eq.(\ref{ansatz-eq}), the presence of a {\it ``random''} fixed point corresponds to 
the case $u(r)\rightarrow - \infty$ and $\hat{\Delta}(r)=\hat{\Delta}^\star$. 
The phase transition occurring at the {\it ``random''} 
fixed point belongs to a new universality class whose exponents are non universal and depend 
upon the value $\hat{\Delta}^\star$. It is possible to compute the
exponents $\nu_{\perp}$ and $\alpha$ by fixing the value of $\hat{\Delta}=
\hat{\Delta}^\star$ in eq.(\ref{RG-1}) and proceeding as in the pure problem. 
One finds that such exponents are given by the same formulas as in the pure case but 
with a renormalized parameter $g_R= g - d\hat{\Delta}^\star$.
A rich scenario arises where the order of the phase transition is changed 
by disorder. The range of parameters where a continuous phase transition occurs 
in the disordered system overlaps the one of continuous phase transitions in the 
pure system in $d=1$, but in higher dimensions regions of the phase space exist where a continuous 
phase transition substitutes an otherwise first order phase transition, thus shifting the upper 
critical dimensionality toward higher values of $g$.\\
Another important effect of disorder is to change the sign of the specific heat exponent.
In $d=1$ and $g \geq 0$, the {\it ``random''} critical exponent of the specific heat is negative, 
unlike that of the pure system: the specific heat is thus finite at the critical point
unlike the divergent specific heat of the corresponding pure case. Similar features are predicted
in higher dimensions where a finite specific heat takes the place of an otherwise divergent one 
in the range of $g$-values where the {\it ``random''} solution exists. 
The critical exponents $\nu_{\perp}$ and $\alpha$ both at the pure and at 
the {\it ``random''} critical point are collected in Table 1.
\begin{table}[H]
\caption{\label{table1}Critical exponents at the {\it pure} (1) and 
{\it random} critical point (2), with $g_R= g - d\hat{\Delta}^\star$.}
\begin{ruledtabular}
\begin{tabular}{lll}
$\;\;$ Exponents&$\;\;\;\;\;\;\;\;\;$Formulas $\;$ \\
\hline
$\;\;$$\nu_{\perp , (1)}^{-1}$&$\;\;\;\sqrt{(d-2)^2+4g}$ \\
$\;\;$$\alpha_{(1)}$& $2\left[ 1- d \left((d-2)^2+4g \right)^{-\frac{1}{2}}\right]$ \\
$\;\;$$\nu_{\perp , (2)}^{-1}$&$\;\;\;\sqrt{(d-2)^2+4g_R}$ \\
$\;\;$$ \alpha_{(2)}$& $2\left[ 1- d \left((d-2)^2+4g_R \right)^{-\frac{1}{2}}\right]$\\
\end{tabular}
\end{ruledtabular}
\end{table}
We conclude discussing the case of $g=0$ and $d=1$, corresponding to the situation where 
no long range interaction is considered.
The RG flow is shown in Fig.\ref{RGcartoon2}: it is evident that the location of the 
{\it ``random''} critical point coincides with that of the pure case, i.e.
$\hat{\Delta}^\star=u^\star=0$.
There are, however, interesting features of the phase transition in the disordered system 
to be stressed.
Although in the continuous limit ($\Lambda\rightarrow+\infty$) $u(0)=0$ and thus the 
introduction of small disorder gives rise to an RG flow toward strong coupling regimes, 
at finite cutoffs and depending on the initial condition $\hat{\Delta}^0$ and $u(0)$,
the RG flux can flow either to the strong coupling regime or toward the {\it ``pure''}
stable fixed point. For $0<u(0)=\hat{\Delta}^0<<1$ a separatrix exits between these two regions 
and along it the RG flux flows toward the critical point $\hat{\Delta}^\star=u^\star=0$.\\ 
To summarize, this paper contains an RG analysis of depinning transitions of two directed heteropolymers
in the presence of an effective long range interaction. The existence of a new {\it ``random''} critical point
has been predicted: two exponents have been exactly computed and found to depend upon disorder. 
Depending on the initial value of the disorder variance $\Delta^0$, the RG flux either flows 
toward strong coupling regime or to the new {\it ``random''} fixed point. The order of the phase transition 
is changed by disorder as well as the sign of the specific heat exponent. Similar results are expected 
to hold in related problems.
\acknowledgments
\begin{acknowledgments}
The author should like to thank Prof.~J.T.~Chalker, Dr.~D.~Marenduzzo and Dr.~R.~Rajesh for helpful conversations
in the early stages of this work.
He is grateful to Prof.~C.~Di Castro, Prof.~C.~Castellani and Prof.~M.~Grilli for valuable discussions and 
important comments.
He is also indebted to Dr.~J.~Lorenzana and Dr.~M.~Capone for critical reading of this manuscript.
Support from the {\it  ``Fondazione Della Riccia''} (Florence, Italy) and a Junior Member affiliation 
from the {\it ``Isaac Newton Institute for Mathematical Sciences''} (Cambridge, UK) are gratefully 
acknowledged.
\end{acknowledgments}

\end{document}